\documentclass[11pt]{article} 
\usepackage{amssymb}
\usepackage{amstext}
\usepackage{latexsym}

\newcommand{\CC}{{\mathcal{C}}}

\newcommand{\E}{{\mathsf{E}}}
\newcommand{\EE}{{\mathcal{E}}}

\newcommand{\G}{{\mathcal{G}}}
\newcommand{\HH}{{\mathcal{H}}}

\newcommand{\V}{{\mathcal{V}}}

\newcommand{\X}{{\mathcal{X}}}
\newcommand{\Y}{{\mathcal{Y}}}
\newcommand{\ZZ}{{\mathcal{Z}}}

\newcommand{\ie}{{\em i.e., }}
\newcommand{\eg}{{\em e.g., }}

\newcommand{\etal}{\emph{et al.\ }}

\newcommand{\inner}[2]{\langle{#1},{#2}\rangle}

\newcommand{\half}{\frac{1}{2}}

\newcommand{\SNR}{\mathrm{SNR}}  %(math mode only)
\newcommand{\openbox}{\leavevmode
  \hbox to.77778em{%
  \hfil\vrule
  \vbox to.675em{\hrule width.6em\vfil\hrule}%
  \vrule\hfil}}
\newcommand{\proofname}{Proof}

\begin{document}
\renewcommand{\textfraction}{0}

\title{Shannon meets Wiener II:  \\
On MMSE estimation in successive decoding schemes}
 \author{\normalsize
G. David Forney, Jr.%\footnote{I am grateful to J. M. Cioffi, U.
%Erez, R. Fischer and R. Zamir for many helpful comments.}
 \\[-5pt]
\small MIT \\[-5pt]Ê
\small Cambridge, MA 02139 USA \\ \small \tt
forneyd@comcast.net }
\date{}
\maketitle
\thispagestyle{empty}
\begin{abstract}
We continue to discuss why MMSE estimation arises in coding schemes that
approach the capacity of linear Gaussian channels.  Here we consider
schemes that involve successive decoding, such as decision-feedback
equalization or successive cancellation.
\end{abstract}

\begin{quote}
``Everything should be made as simple as possible, but not simpler."--- A.
Einstein.
\end{quote}

\section{Introduction}

The occurrence of minimum-mean-squared-error (MMSE) linear estimation
filters in constructive coding schemes that approach information-theoretic
limits of linear Gaussian channels has been repeatedly observed, and
justified by various arguments.  For example, in an earlier paper \cite{F03}
we showed the necessity of the MMSE estimation factor in the
capacity-approaching lattice coding scheme of Erez and Zamir \cite{EZ01} for
the classic additive white Gaussian noise (AWGN) channel.

In particular, MMSE decision-feedback equalizer (MMSE-DFE) filters have been
used in 
\linebreak coding schemes that approach the capacity of linear Gaussian
intersymbol interference (ISI) channels \cite{CDEFI95}, and
generalized MMSE-DFE (MMSE-GDFE) filters have been used in coding schemes
that approach the capacity region of multiple-input, multiple-output (MIMO)
linear Gaussian channels \cite{CF97}.  These successive decoding schemes
combine ``analog" discrete-time linear MMSE estimation with the essentially
``digital" assumption of ideal decision feedback (perfect
prior decisions).

The fact that MMSE filters allow information-theoretic limits to be
approached in successive decoding scenarios is widely understood, and has been
proved in various ways.  Our aim here is to provide the simplest and most
transparent justification possible.  Some principal features of our
approach are:
\begin{itemize}
\item As in \cite{CF97, GV02}, we use a geometric Hilbert space formulation;
\item Our results are  based mainly on the sufficiency property of MMSE
estimators, with \\ information-theoretic results mostly as corollaries;
\item Proofs of almost all results are given. All
proofs are brief and straightforward.
\end{itemize} 

In developing this approach, we have benefited from our earlier work with
Cioffi \etal \cite{CDEFI95, CF97} and from the insightful development of Guess
and Varanasi \cite{GV98, GV02}.  We would also like to acknowledge helpful
comments on earlier drafts of this paper by G. Caire, J. Cioffi, U. Erez, T.
Guess, S. Shamai and G. Wornell.

\subsection{Hilbert spaces of jointly Gaussian random variables}

All random variables in this note will be finite-variance, zero-mean, proper
(circularly symmetric) complex Gaussian random variables.  Random variables
will be denoted by capital letters such as $X$.  If the variance $\sigma^2$
of $X$ is nonzero, then $X$ has a probability density function
(pdf) 
$$
p_X(x) = \frac{1}{\pi \sigma^2} \exp - \frac{|x|^2}{\sigma^2},
$$
and thus its differential entropy is 
$
h(X) = \E[-\log p_X(x)] = \log  \pi e \sigma^2.
$
If the variance of $X$ is zero, then $X$ is the deterministic zero variable
0. 

Sets of such random variables will be denoted by a script letter such as $\X =
\{X_i\}$.  In this paper, we will consider only finite sets of random
variables.  A particular application may involve a finite set of such sets
such as $\{\X, \Y,\ZZ\}$.  

Whenever we have a set of Gaussian variables, their statistics will be
assumed to be jointly Gaussian.  A set of variables is jointly Gaussian if
they can all be expressed as linear combinations of a common set of
independent Gaussian random variables.  It follows that any set of  linear
combinations of jointly Gaussian random variables is jointly Gaussian.

The set of all complex linear combinations of a given finite set $\X$ of
finite-variance, zero-mean, proper jointly Gaussian complex random variables
is evidently a complex vector space $\G$.  Every element of $\G$ is a
finite-variance, zero-mean, proper complex Gaussian random variable, and
every subset of $\G$ is jointly Gaussian. The zero vector of $\G$ is the
unique zero variable 0.  The dimension of $\G$ is at most the size $|\X|$
of $\X$.

It is well known that if an inner product is defined on $\G$ as the
cross-correlation $\inner{X}{Y} = \E[XY^*]$, then $\G$ becomes a Hilbert
space (a complete inner product space), a subspace of the Hilbert space $\HH$
consisting of all finite-variance zero-mean complex random
variables.  The squared norm of $X \in \G$ is then its
variance, $||X||^2 = \inner{X}{X} = \E[|X|^2]$.  Variances are real, finite
and strictly non-negative;  \ie if $X \in \G$ has zero
variance, $||X||^2 = 0$, then $X$ must be the deterministic zero variable, $X
= 0$.

If $\G$ is generated by $\X$, then all inner products between elements of
$\G$ are determined by the inner product (autocorrelation) matrix $R_{xx} =
\{\inner{X}{X'} \mid X,X' \in \X\}$ (the Gram matrix of $\X$). 
In other words, the matrix $R_{xx}$ completely determines the geometry of
$\G$.  Since all subsets of variables in $\G$ are jointly Gaussian, the joint
statistics of any such subset of $\G$ are completely determined by their
second-order statistics, and thus by $R_{xx}$.  

A subset $\Y \subset \G$ is called \emph{linearly dependent} if there is some
linear combination of the elements of $\Y$ that is equal to the zero variable
0, and linearly independent otherwise.  We will see that a subset $\Y \subset
\G$ is linearly independent if and only if its autocorrelation matrix $R_{yy}$
has full rank.

Two random variables are \emph{orthogonal} if their inner product
is zero;  \ie if they are uncorrelated.  If two jointly Gaussian variables
are orthogonal, then they are statistically independent.  The only variable
in $\G$ that is orthogonal to itself (\ie satisfies $\inner{X}{X} = 0$) is
the zero variable 0.  If
$\inner{X}{Y} = 0$, then the \emph{Pythagorean theorem} holds:  $$||X + Y||^2
= ||X||^2 + ||Y||^2.$$ 

Given any subset $\Y \subset \G$, the \emph{closure}
$\overline{\Y}$ of $\Y$, or the \emph{subspace generated by} $\Y$, is the set
of all linear combinations of elements of $\Y$.
Also, the set of all $X \in \G$ that are orthogonal to all elements of
$\Y$ is a subspace of $\G$, called the \emph{orthogonal subspace}
$\Y^\perp \subseteq \G$.  Since 0 is the only element of $\G$ that is
orthogonal to itself, the only common element of $\overline{\Y}$ and
$\Y^\perp$ is
$0$.

\subsection{The projection theorem}

The key geometric property of the Hilbert space $\G$ is 
the \textbf{projection theorem}: if $\V$ and $\V^\perp$ are orthogonal
subspaces of
$\G$, then there exists a \emph{unique} $X_{|\V} \in \V$ and $X_{\perp\V}
\in \V^\perp$ such that $X = X_{|\V} + X_{\perp\V}$.  $X_{|\V}$ and
$X_{\perp\V}$ are called the \emph{projections} of $X$ onto $\V$ and
$\V^\perp$, respectively.

A explicit formula for a projection $X_{|\V}$ such that $X - X_{|\V} \in
\V^\perp$ will be given below.  Uniqueness is the most important part of the
projection theorem, and may be proved as follows:  if $X = Y + Z$ and also $X
= Y' + Z'$, where
$Y, Y'
\in \V$ and $Z, Z' \in \V^\perp$, then
$$
0 = ||X - X||^2 = ||Y - Y'||^2 + ||Z - Z'||^2,
$$
where the Pythagorean theorem applies since $Y - Y' \in \V$ and $Z - Z'
\in \V^\perp$.  Since norms are non-negative, this implies $||Y -
Y'||^2 = ||Z - Z'||^2 = 0$, which implies $Y = Y'$ and $Z = Z'$.

The projection theorem is illustrated by the little ``Pythagorean" diagram
below.  Since $X_{|\V}$ and $X_{\perp\V}$ are orthogonal, we have 
$||X||^2 = ||X_{|\V}||^2 + ||X_{\perp\V}||^2.$

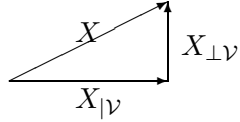
\setlength{\unitlength}{5pt}
\begin{center}
\begin{picture}(16,6)(0,1)
\put(0,2){\vector(1,0){12}}
\put(5,0){$X_{|\V}$}
\put(12,2){\vector(0,1){6}}
\put(13,4){$X_{\perp\V}$}
\put(0,2){\vector(2,1){12}}
\put(5,5){$X$}
\end{picture}
\end{center}

%The projection theorem defines linear orthogonal self-adjoint projection
%operators $P_{\V}: \G \to \G$ and $P_{\V^\perp}: \G \to \G$ such that
%$P_{\V}(X) = X_{|\V}$ and $P_{\V^\perp}(X) = X_{\perp\V}$.  These 
%operators satisfy the projection identities
%$P_{\V} + P_{\V^\perp} = I$,
%$(P_{\V})^2 = P_{\V}$ and $(P_{\V^\perp})^2 = P_{\V^\perp}$, and
%$P_{\V}P_{\V^\perp} = P_{\V^\perp}P_{\V} = 0$.  

If $\overline{\Y}$ is a subspace that is generated by a set of variables $\Y$,
then with  mild abuse of notation we will write $X_{|\Y}$ and $X_{\perp\Y}$
rather than
$X_{|\overline{\Y}}$ and
$X_{\perp\overline{\Y}}$.
%, and similarly $P_{\Y}$ and
%$P_{\Y^\perp}$ rather than $P_{\overline{\Y}}$ and
%$P_{\overline{\Y}^\perp}$.

\subsection{Innovations representations}

Let $\X \subset \G$ be a finite subset of elements of $\G$,
and let $\overline{\X} \subseteq \G$ be the subspace of $\G$ generated by
$\X$.  An orthogonal basis for $\overline{\X}$ may then be found by a
recursive (Gram-Schmidt) decomposition, as follows.

Denote the elements of the generator set $\X$ by $X_1, X_2, \ldots$, and let
$\overline{\X_1^{i-1}}$ denote the subspace of $\G$ generated by
$\X_1^{i-1} = \{X_1, X_2, \ldots, X_{i-1}\}$.  To initialize, set $i = 1$ and
$\X_1^{0} =
\emptyset$.  For the $i$th recursion, using the projection theorem, write
$X_i$ uniquely as
$$
X_i = (X_i)_{|\X_1^{i-1}} + (X_i)_{\perp\X_1^{i-1}}.
$$
We have $(X_i)_{\perp\X_1^{i-1}} = 0$ if and only if $X_i \in
\overline{\X_1^{i-1}}$.  In this case
$\overline{\X_1^{i}} = \overline{\X_1^{i-1}}$, so we can delete $X_i$ from
the generator set $\X$ without affecting $\overline{\X}$.  Otherwise, we can
take the ``innovation" variable $E_i = (X_i)_{\perp\X_1^{i-1}} \neq 0$
as a replacement for $X_i$ in the generator set;  the
space generated by $\X_1^{i-1}$ and $E_{i}$ is still
$\overline{\X_1^{i}}$, but
$E_i$ is orthogonal to $\overline{\X_1^{i-1}}$.  By induction, the nonzero
innovations variables up to $E_i$ generate $\overline{\X_1^{i}}$ and are
mutually orthogonal;  \ie they form an orthogonal basis for
$\overline{\X_1^{i}}$.

This recursive decomposition thus shows that:
\begin{itemize}
\item Any generator set $\X$ for a subspace $\overline{\X}$ contains a
linearly independent generator set $\X' \subseteq \X$ that generates
$\overline{\X}$.  Therefore, without loss of generality, we may assume that
any generator set $\X$ for $\overline{\X}$ is linearly independent.
\pagebreak
\item  Given a linearly independent generator set $\X = \{X_1, X_2, \ldots\}$
for $\overline{\X}$, we can find an orthogonal basis $\EE = \{E_1,
E_2, \ldots\}$ for $\overline{\X}$, where $E_i = (X_i)_{\perp\X_1^{i-1}} =
X_i - (X_i)_{|\X_1^{i-1}}$.
Since $(X_i)_{|\X_1^{i-1}}$ is a linear combination of $X_1, X_2, \ldots,
X_{i-1}$, if we write $\X$ and $\EE$ as column vectors, then we have
$$
\EE = L^{-1}\X,
$$
where $L^{-1}$ is a monic (\ie having ones on the diagonal) lower triangular
matrix.
Since $L^{-1}$ is square and has a monic lower triangular inverse
$L$, we may write alternatively
$$
\X = L\EE.
$$
\end{itemize}

We conclude that a finite set of random variables $\X$ is
jointly Gaussian if and only if $\X$ can be written as a monic lower
triangular (``causal") linear transformation $\X = L\EE$ of an orthogonal
innovations sequence $\EE$.  All innovations variables are nonzero (\ie $\EE$
is linearly independent) if and only if $\X$ is linearly independent.  This is
called an \emph{innovations representation} of $\X$. 

Moreover, the expression $\X = L\EE$ implies that the autocorrelation
matrix of $\X$ is
$$
R_{xx} = L R_{ee} L^* = L D^2 L^*,
$$
where $L^*$ denotes the conjugate transpose of $L$ (a monic upper triangular
matrix), and $R_{ee}$  is a non-negative real diagonal matrix $D^2$, because
$\EE$ is an orthogonal sequence.  This is called a
\emph{Cholesky decomposition} of  $R_{xx}$;  the diagonal elements
$||E_i||^2$ of
$D^2$ are called the \emph{Cholesky factors} of $R_{xx}$.  The Cholesky
factors are all nonzero, and thus $R_{ee}$ and $R_{xx}$ have full rank, if
and only if $\X$ is linearly independent. In general,
the rank of $R_{xx}$ is the number of nonzero innovations variables $E_i$
in the innovations representation $\X = L\EE$.

Since $L$ is monic lower triangular, its determinant is 1: $|L| = |L^*| =
1$.  Therefore
$$
|R_{xx}| = |R_{ee}| = \prod_i ||E_i||^2.
$$

\subsection{Differential entropy}

To find the differential entropy $h(\X)$ of a set $\X$
of $N$ linearly independent jointly Gaussian random variables, we first
recall that the differential entropy of a complex Gaussian variable
$X$ with variance $||X||^2 > 0$ is
$h(X) = \log  \pi e ||X||^2.$
Then we have
\begin{eqnarray*}
h(\X) & = & h(X_1) + h(X_2 \mid X_1) + \cdots h(X_i \mid \X_1^{i-1}) + \cdots
\\ & = & h(E_1) + h(E_2) + \cdots h(E_i) + \cdots \\
& = & \log  \pi e ||E_1||^2 + \log  \pi e ||E_2||^2 + \cdots + \log  \pi e
||E_i||^2 + \cdots \\
& = & \log (\pi e)^N |R_{ee}| \\
& = & \log (\pi e)^N |R_{xx}|,
\end{eqnarray*}
where we use the chain rule of differential entropy, we note that $E_i = X_i
- (X_i)_{|\X_1^{i-1}}$ implies $h(E_i) = h(X_i \mid \X_1^{i-1})$, and we
apply the determinantal equalities that arise from the innovations
representation of
$\X$.

Thus the differential
entropy per complex dimension is 
$$\frac{h(\X)}{N} = \log \pi e|R_{xx}|^{1/N},$$
where $|R_{xx}|^{1/N}$ is the geometric mean of the Cholesky factors (or
eigenvalues) of $R_{xx}$. 
%Thus the volume of the typical sequence region $T_\X$ of $\X$ is of the
%exponential order of $\det R_{xx}$:
%$$
%V(T_\X) \approx (2\pi e)^N \det R_{xx}.
%$$
Note that this result is independent of the order in which we take the
variables in $\X$.

\pagebreak
\subsection{Fundamentals of MMSE estimation theory}

Suppose that $X$ represents a random variable to be estimated and
that $\Y$ represents a set of observed variables, where $X$ and $\Y$ are
jointly Gaussian.  A \emph{linear estimate} of $X$ is a linear function
of
$\Y$; \ie a random variable $V$ in the space $\overline{\Y}$.  The
\emph{estimation error} is then $E = X - V$. 

By the projection theorem, the projection $X_{|\Y} \in \overline{\Y}$
minimizes  the estimation error variance
$||X-V||^2$ over $V \in \overline{\Y}$, because, using the Pythagorean
theorem and the fact that $X_{|\Y} - V \in \overline{\Y}$ while $X_{\perp\Y}
\in \overline{\Y}^\perp$, we have
$$
||X - V||^2 = ||X_{|\Y} + X_{\perp\Y} - V||^2 = ||X_{|\Y} - V||^2 +
||X_{\perp\Y}||^2 \ge ||X_{\perp\Y}||^2,
$$
with equality if and only if $V = X_{|\Y}$.  For this reason $X_{|\Y}$
is called the \emph{minimum-mean-squared error} (MMSE) \emph{linear estimate}
of $X$ given $\Y$, and $X_{\perp\Y}$ is called the \emph{MMSE estimation
error}.  Moreover, the \emph{orthogonality principle} holds:  $V \in
\overline{\Y}$ is the MMSE linear estimate of $X$ given $\Y$ if and only if
$X - V$ is orthogonal to $\overline{\Y}$.

Similarly, if $\X \subseteq \G$ is a set of random variables, then by the
orthogonality principle the set
$\V \in \overline{\Y}$ is the corresponding set $\X_{|\Y}$ of MMSE linear
estimates of $\X$ given $\Y$ if and only if $\inner{\X - \V}{\Y} = 0$, or
$\inner{\V}{\Y} = \inner{\X}{\Y}$.  Writing $\V$ as a set of
linear combinations of the elements of $\Y$ in matrix form, namely $\V =
A_{xy} \Y$, and defining $R_{xy}$ as the
cross-correlation matrix $\inner{\X}{\Y}$ and $R_{yy}$ as the
autocorrelation matrix $\inner{\Y}{\Y}$, we obtain a unique solution
$$
A_{xy} = R_{xy} R_{yy}^{-1},
$$
where without loss of generality we assume that $R_{yy}$ is
invertible; \ie that $\Y$ is a linearly independent generator set for
$\overline{\Y}$.  In short, an explicit formula for the projection of $\X$
onto $\overline{\Y}$ is  $$\X_{|\Y} = R_{xy} R_{yy}^{-1}\Y.$$

The expression $\X = A_{xy}\Y + \X_{\perp\Y}$ shows that $\X$ may
be regarded as the sum of a linear estimate derived from $\Y$ and an
independent error (innovations) variable $\EE = \X_{\perp\Y}$.  This
decomposition is illustrated in the block diagram below.

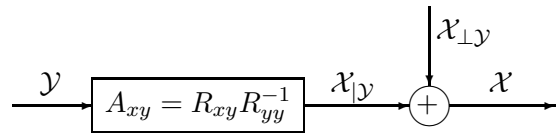
\setlength{\unitlength}{5pt}
\begin{center}
\begin{picture}(41,8)
\put(0,2){\vector(1,0){6}}
\put(2,3){$\Y$}
\put(6,0){\framebox(16,4){$A_{xy} = R_{xy} R_{yy}^{-1}$}}
\put(24,3){$\X_{|\Y}$}
\put(22,2){\vector(1,0){8}}
\put(31.5,2){\circle{3}}
\put(30.5,1.5){$+$}
\put(33,2){\vector(1,0){8}}
\put(36,3){$\X$}
\put(31.5,9.5){\vector(0,-1){6}}
\put(32,7){$\X_{\perp\Y}$}
\end{picture}
\end{center}

Since $\X_{\perp\Y}$ has zero mean and is independent of $\Y$, we have
$\E[\X\mid \Y] =\X_{|\Y}$;  \ie the MMSE linear estimate $\X_{|\Y}$ is the
\emph{conditional mean} of $\X$ given $\Y$.  Indeed, this
decomposition shows that the conditional distribution of
$\X$ given $\Y$ is Gaussian with mean $\X_{|\Y}$ and autocorrelation matrix
$R_{ee} = R_{xx} - R_{xy}R_{yy}^{-1}R_{yx}$, by Pythagoras.  Thus
$\X_{|\Y}$ is evidently the
\emph{unconstrained} MMSE estimate of $\X$ given $\Y$; \ie our earlier
restriction to a linear estimate is no real restriction.

Moreover, this block diagram implies that the MMSE estimate $\X_{|\Y}$
is a \emph{sufficient statistic} for estimation of $\X$ from
$\Y$, since $\Y - \X_{|\Y} - \X$ is evidently
 a Markov chain;  \ie $\Y$ and $\X$ are conditionally independent given
$\X_{|\Y}$.  We call this the
\textbf{sufficiency property} of the MMSE estimate. This implies that
$\X$ can be estimated as well from the projection $\X_{|\Y}$ as from $\Y$, so
there is no loss of estimation optimality if we first reduce $\Y$ to
$\X_{|\Y}$. 

Actually, $\X_{|\Y}$ is a \emph{minimal} sufficient
statistic; \ie $\X_{|\Y}$ is a function of every other sufficient statistic
$f(\Y)$. This follows from the fact that the conditional distribution of $\X$
given $f(\Y)$ must be the same as the conditional distribution given $\Y$,
which implies that the conditional mean $\X_{|\Y}$ can be determined from
$f(\Y)$.

\subsection{A bit of information theory}

By the sufficiency property, the MMSE estimate $\X_{|\Y}$ is a function of
$\Y$ that satisfies the data processing inequality of information theory with
equality:  $I(\X; \Y) = I(\X; \X_{|\Y})$.  In other words, the reduction of
$\Y$ to  $\X_{|\Y}$ is \emph{information-lossless}.

Moreover, since $\X = A_{xy}\Y +
\X_{\perp\Y}$ is a linear Gaussian channel model with Gaussian input $\Y$,
Gaussian output $\X$, and independent additive Gaussian noise
$\EE = \X_{\perp\Y}$, we have
$$
I(\X; \Y) = h(\X) - h(\X\mid\Y) = h(\X) - h(\EE) = \log
\frac{|R_{xx}|}{|R_{ee}|},
$$
where we recall that the differential entropy of a set $\X$ of $N$
complex Gaussian random variables with nonsingular autocorrelation matrix
$R_{xx}$ is  $h(\X) =
\log (\pi e)^N |R_{xx}|$.  (We assume that $R_{ee}$ is nonsingular, else
$\{\X, \Y\}$ is linearly dependent, so at least one dimension of $\X$ may be
determined precisely from $\Y$ and $I(\X;\Y) = \infty$.)  

%Since the autocorrelation matrix of $\X_{|\Y}$ is $R_{xy}R_{yy}^{-1}R_{yx}$,
%by Pythagoras we have explicitly $R_{ee} = R_{xx} -
%R_{xy}R_{yy}^{-1}R_{yx}$.  
%(When $\X$ and $\Y$ are one-dimensional, this
%reads $||E||^2 = ||X||^2 - |\inner{X}{Y}|^2/||Y||^2$.  Since $||E||^2 \ge 0$,
%this proves the Schwartz inequality,
%$|\inner{X}{Y}|^2 \le ||X||^2||Y||^2$, with equality if and only if $X$ and
%$Y$ are linearly dependent.)

\subsection{Chain rule of MMSE estimation}\label{SMMSE}

Suppose that $\X, \Y, \ZZ$ are jointly Gaussian sets of random variables and
that we wish to estimate $\X$ based on $\Y$ and $\ZZ$.  The MMSE estimate is
then $\X_{|\Y\ZZ}$, the projection of $\X$ onto the subspace $\overline{\Y} +
\overline{\ZZ}$ generated by the variables in both $\Y$ and $\ZZ$.

The subspace $\overline{\Y} + \overline{\ZZ}$ may be written as the sum of
two orthogonal subspaces as follows:
$$
\overline{\Y} + \overline{\ZZ} = \overline{\Y} + \left(
\overline{\Y}^\perp \cap \overline{\ZZ_{\perp\Y}}\right).
$$
Correspondingly, we may write the projection $\X_{|\Y\ZZ}$ as the sum of two
orthogonal projections as follows:
$$
\X_{|\Y\ZZ} = \X_{|\Y} + (\X_{\perp\Y})_{|\ZZ_{\perp\Y}}.
$$
We call this the \emph{chain rule of MMSE estimation}.  It is illustrated
below:

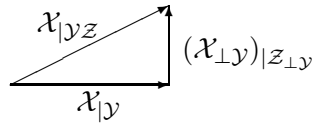
\setlength{\unitlength}{5pt}
\begin{center}
\begin{picture}(16,6)(0,1)
\put(0,2){\vector(1,0){12}}
\put(5,0){$\X_{|\Y}$}
\put(12,2){\vector(0,1){6}}
\put(13,4){$(\X_{\perp\Y})_{|\ZZ_{\perp\Y}}$}
\put(0,2){\vector(2,1){12}}
\put(2,6){$\X_{|\Y\ZZ}$}
\end{picture}
\end{center}

Generalizing, if we wish to estimate $\X$ based on a sequence $\Y = \{\Y_1,
\Y_2, \ldots\}$ of random variables such that $\X$ and $\Y$ are jointly
Gaussian, then the chain rule of MMSE estimation becomes
$$
\X_{|\Y} = \X_{|\Y_1} + (\X_{\perp\Y_1})_{|(\Y_2)_{\perp\Y_1}} + \cdots +
(\X_{\perp\Y_1^{i-1}})_{|(\Y_i)_{\perp \Y_1^{i-1}}} + \cdots,
$$
where $\Y_1^{i-1} = \{\Y_1, \Y_2, \ldots, \Y_{i-1}\}$.
The incremental estimate $(\X_{\perp\Y_1^{i-1}})_{|(\Y_i)_{\perp \Y_1^{i-1}}}$
thus represents the ``new information" given by the innovations component
$(\Y_i)_{\perp \Y_1^{i-1}}$ of the observation
$\Y_i$ about
$\X$, given the previous observations $\Y_1^{i-1}$.

The innovations representation may be seen as a special case of the chain rule
of MMSE estimation.  Indeed, if $\X = \{X_1, X_2, \ldots\}$ and we take $\Y =
\X$, then $\X_{|\X} = \X$, and the ``new information" sequence becomes 
$$
(\X_{\perp\X_1^{i-1}})_{|(X_i)_{\perp \X_1^{i-1}}}
=(\X_{\perp\X_1^{i-1}})_{|E_i} =  \{0, \ldots, 0,
E_i,
\ldots\};
$$
\ie the  first $i$ components of $(\X_{\perp \X_1^{i-1}})_{|X_i}$ are
$\{0, \ldots, 0, E_i\}$,
where $E_i = (X_i)_{\perp\X_1^{i-1}}$ is the $i$th innovation variable of
$\X$; the remaining components are evidently linearly
dependent on $E_i$.

\section{Successive decoding}

Often it is natural or helpful to regard a set $\X$ of Gaussian random
variables as a sequence of subsets, $\X = \{\X_1, \X_2, \ldots\}$.  For
instance $\X_1, \X_2, \ldots$ might represent a discrete-time sequence, in
which case the ordering naturally follows the time ordering;  or, in a
multi-user scenario, $\X_1, \X_2, \ldots$ might represent different users, in
which case the ordering may be arbitrary.  Thus the
index set $\{1, 2, \ldots\}$ indicates an ordering, but is not
necessarily a time index set.

Our aim will be to signal at a rate approaching the mutual information $I(\X;
\Y)$.  As above, we may write
$$
I(\X; \Y) = I(\X; \X_{|\Y}) = h(\X) - h(\EE) = \log \frac{|R_{xx}|}{|R_{ee}|},
$$
where $\EE = \{\EE_1, \EE_2, \ldots\}$ is the sequence of estimation error
subsets $\EE_i = (\X_i)_{\perp \Y}$.

We will consider a successive decoding scenario in which the
subsets $\X_1, \X_2, \ldots$ are detected sequentially from a set $\Y$ of
observed variables.  For each index $i$, we will aim to signal at a rate
approaching the incremental rate
$$
R_i = h(\X_i \mid \X_i^{i-1}) - h(\EE_i \mid \EE_i^{i-1}),
$$
where $\X_i^{i-1} = \{\X_1, \X_2, \ldots, \X_{i-1}\}$ and $\EE_i^{i-1} =
(\X_i^{i-1})_{\perp \Y}$.  By the chain rule of differential entropy, we will
then approach a total rate of $\sum_i R_i = h(\X) - h(\EE) = I(\X;
\Y)$.

 For successive decoding, we will make the following critical assumption:

\begin{quote}
\textbf{Ideal decision feedback assumption}:  In the detection of the
variable subset $\X_i$, the values of the previous variables
$\X_i^{i-1}$ are known precisely.
\end{quote}

The ideal decision feedback assumption is the decisive break between the
classical analog estimation theory of Wiener \etal and the digital Shannon
theory.  If the $\X_i$ are continuous Gaussian variables, then in
general it is nonsense to suppose that they can be estimated precisely
(assuming that $\X$ and $\Y$ are not linearly dependent).  On the
other hand, if the
$\X_i$ are codewords in some discrete code $\CC$ whose words are chosen
randomly according to the Gaussian statistics of $\X_i$ given $\X_i^{i-1}$,
and if the length of $\CC$ is large enough and the rate of $\CC$ is less than
the incremental rate $R_i$, then Shannon theory
shows that the probability of not decoding $\X_i$ precisely given $\Y$ and
$\X_1^{i-1}$ may be driven arbitrarily close to 0.  So in a digital
coding scenario, the ideal decision feedback assumption may be quite
reasonable.  
%(Later, we will discuss additional coding scenarios in which it may hold.)

The MMSE estimate $(\X_i)_{|\Y, \X_1^{i-1}}$ of
$\X_i$ is a sufficient statistic for estimation of $\X_i$ given
$\Y$ and $\X_1^{i-1}$.
Moreover, by the chain rule of MMSE estimation, we may alternatively write
$$
(\X_i)_{|\Y, \X_1^{i-1}} =
 (\X_i)_{|\Y} + ((\X_i)_{\perp\Y})_{|(\X_1^{i-1})_{\perp\Y}} =
 (\X_i)_{|\Y} + (\EE_i)_{|\EE_1^{i-1}}.
$$
The estimation error is
$(\EE_i)_{\perp\EE_1^{i-1}}$.  In short,
$\X_i$ is the sum of three independent components: the MMSE estimate of $\X_i$
given $\Y$, the MMSE prediction of $\EE_i$ given $\EE_1^{i-1}$, and the
estimation error $(\EE_i)_{\perp\EE_1^{i-1}}$.
The differential entropy of the estimation
error may thus be written in any of the following ways:
$$
h((\EE_i)_{\perp \EE_i^{i-1}}) = h(\X_i \mid \Y, \X_i^{i-1}) = 
 h(\EE_i \mid \EE_i^{i-1}).
$$

We note therefore that $\sum_i R_i = I(\X; \Y)$ follows alternatively from 
the chain rule of mutual information, since
$$
I(\X_i; \Y \mid \X_i^{i-1}) = h(\X_i \mid \X_i^{i-1}) - h(\X_i \mid \Y,
\X_i^{i-1}) = h(\X_i \mid \X_i^{i-1}) - h(\EE_i \mid \EE_i^{i-1}) = R_i.
$$

Successive decoding then works as follows.  The sequence to be decoded is
$\X_1, \X_2, \ldots$, and the observed sequence is $\Y$.  We first reduce $\Y$
to the MMSE estimate $(\X_1)_{|\Y}$ and decode $\X_1$ from it, in the presence
of the error $\EE_1 =  (\X_1)_{\perp\Y}$.  If the decoding of $\X_1$ is
correct, then we can compute $\EE_1 = \X_1 - (\X_1)_{|\Y}$ and form the
estimate
$(\EE_2)_{|\EE_1}$, which we add to $(\X_2)_{|\Y}$ to form the input to a
decoder for $\X_2$ with error $(\EE_2)_{\perp\EE_1}$, and so forth. 

This ``decision feedback" scheme is illustrated in the figure below.  The
``forward filter" $A_{xy}$ is the MMSE estimator of the sequence $\X$ given
$\Y$.  The ``backward filter" is the MMSE predictor of
$\EE_i$ given $\EE_1^{i-1}$, where ideal decision feedback is
assumed in computing the previous error $\EE_1^{i-1}$.

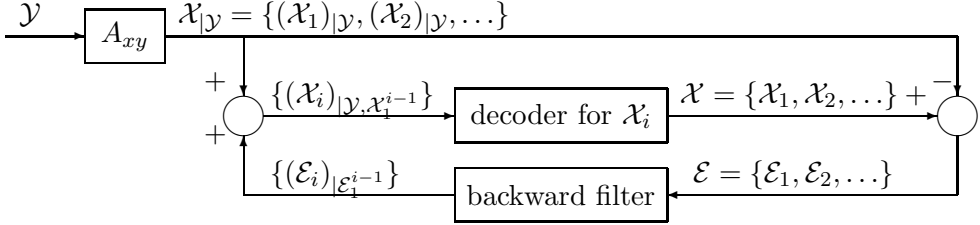
\setlength{\unitlength}{5pt}
\begin{center}
\begin{picture}(74,14)(0,2)
\put(0,14){\vector(1,0){6}}
\put(1,15){$\Y$}
\put(6,12){\framebox(6,4){$A_{xy}$}}
\put(12,14){\line(1,0){60}}
\put(72,14){\vector(0,-1){4.5}}
\put(13,15){$\X_{|\Y} = \{(\X_1)_{|\Y}, (\X_2)_{|\Y}, \ldots\}$}
\put(18,14){\vector(0,-1){4.5}}
\put(15,10){+}
\put(15,6){+}
\put(18,8){\circle{3}}
\put(19.5,8){\vector(1,0){14.5}}
\put(20,9){$\{(\X_i)_{|\Y,\X_1^{i-1}}\}$}
\put(34,6){\framebox(16,4){decoder for $\X_i$}}
\put(50,8){\vector(1,0){20.5}}
\put(51,9){$\X = \{\X_1, \X_2, \ldots\}$}
\put(70,10){$-$}
\put(68,9){+}
\put(72,8){\circle{3}}
\put(72,6.5){\line(0,-1){4.5}}
\put(72,2){\vector(-1,0){22}}
\put(52,3){$\EE = \{\EE_1, \EE_2, \ldots\}$}
\put(34,0){\framebox(16,4){backward filter}}
\put(34,2){\line(-1,0){16}}
\put(18,2){\vector(0,1){4.5}}
\put(20,3){$\{(\EE_i)_{|\EE_1^{i-1}}\}$}
\end{picture}
\end{center}

This decision-feedback scheme is said to be in ``noise-predictive" form,
since the error sequence $\EE$ is predicted by the causal backward filter. 
By linearity, we can put it into more standard decision-feedback form as shown
below, where the backward filter is denoted by $A_b$:

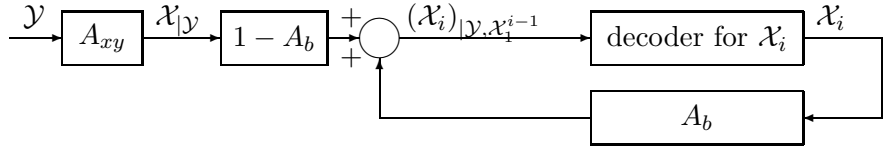
\setlength{\unitlength}{5pt}
\begin{center}
\begin{picture}(56,8)(-5,2)
\put(-10,8){\vector(1,0){4}}
\put(-9,9){$\Y$}
\put(-6,6){\framebox(6,4){$A_{xy}$}}
\put(0,8){\vector(1,0){6}}
\put(1,9){$\X_{|\Y}$}
\put(6,6){\framebox(8,4){$1 - A_{b}$}}
\put(14,8){\vector(1,0){2.5}}
\put(15,9){+}
\put(15,6){+}
\put(18,8){\circle{3}}
\put(19.5,8){\vector(1,0){14.5}}
\put(20,9){$(\X_i)_{|\Y,\X_1^{i-1}}$}
\put(34,6){\framebox(16,4){decoder for $\X_i$}}
\put(50,8){\line(1,0){6}}
\put(51,9){$\X_i$}
\put(56,8){\line(0,-1){6}}
\put(56,2){\vector(-1,0){6}}
\put(34,0){\framebox(16,4){$A_b$}}
\put(34,2){\line(-1,0){16}}
\put(18,2){\vector(0,1){4.5}}
\end{picture}
\end{center}

Successive decoding thus
breaks the joint detection of $\X = \{\X_1, \X_2, \ldots\}$ into a series of
``per-user" steps.  This idea underlies classical decision-feedback schemes
for sequential transmission on a single channel, and also successive
interference cancellation schemes on multi-access channels.

Moreover, if we can achieve a small error probability with a
code of  rate close to $R_i$ for each $i$, then
we can achieve an aggregate rate close to $I(\X; \Y)$ with an error
probability no greater than the sum of the component error probabilities, by
the union bound.  Again, this holds regardless of the ordering of the
users.

In practice, achieving a rate approaching the mutual information will
require very long codes.  This is usually not an obstacle in a multi-access
scenario.  In the case of sequential transmission on a single channel which
is not memoryless, it can be achieved in principle by interleaving beyond the
memory length of the channel (for details, see \cite{GV00}).  
Alternatively, if the channel is
known at the transmitter, then interference may be effectively removed at the
transmitter by various precoding or precancellation schemes (\eg
\cite{CDEFI95, EF92, SL96}).

These schemes naturally extend to infinite jointly stationary and jointly
Gaussian sequences $\X = \{\ldots, \X_0, \X_1, \ldots\}$ and $\Y = \{\ldots,
\Y_0, \Y_1, \ldots\}$.  The forward and backward
filters shown above become time-invariant in the limit.  Cholesky
decompositions become multivariate spectral factorizations.  Sequence mutual
information quantities such as $I(\X; \Y)$ are replaced by information
rates.  For a full development, see Guess and Varanasi \cite{GV02}.  The point
is that the conceptual basis of the development is essentially the same.

In summary, when the signal to be detected and the observation are jointly
Gaussian, and our objective is to maximize mutual information, we may always
incorporate an MMSE estimator into the receiver, because an MMSE estimator is
a sufficient statistic and thus information-lossless.

\footnotesize{

}
\end{document}